\documentclass[conference]{IEEEtran}
\usepackage{times}
\usepackage{blindtext, rotating}

\usepackage{epsfig}
\usepackage{graphicx}
\usepackage{amsmath}
\usepackage{amssymb}

\usepackage{tabularx}

\usepackage[utf8]{inputenc}
\usepackage[T1]{fontenc}
\usepackage{textcomp}
\usepackage{gensymb}
\usepackage{multirow}
\usepackage{booktabs}
\usepackage{xcolor}
\usepackage{subcaption}
\usepackage{float}
\usepackage{capt-of}
\usepackage{etoolbox}
\usepackage{adjustbox}
\usepackage[font=small]{caption}
\usepackage{paralist}
\usepackage{algorithm}
\usepackage{algpseudocode} 
\usepackage{makecell} 

\usepackage{hyperref}

\DeclareMathOperator*{\argmin}{\arg\!\min}

\newcommand{\bo}[1]{\textbf{#1}}

\IEEEoverridecommandlockouts
\def\BibTeX{{\rm B\kern-.05em{\sc i\kern-.025em b}\kern-.08em
    T\kern-.1667em\lower.7ex\hbox{E}\kern-.125emX}}
\begin{document}

\title{Scenic Routes over Points in 2D Space}

\author{\IEEEauthorblockN{Loay Rashid}
\IEEEauthorblockA{\textit{DSAC} \\
\textit{IIIT-Hyderbad}\\
Hyderabad, India \\
loay.rashid@students.iiit.ac.in}
\and
\IEEEauthorblockN{Lini Thomas}
\IEEEauthorblockA{\textit{DSAC} \\
\textit{IIIT-Hyderbad}\\
Hyderabad, India \\
lini.thomas@iiit.ac.in}
\and
\IEEEauthorblockN{Kamalakar Karlapalem}
\IEEEauthorblockA{\textit{DSAC} \\
\textit{IIIT-Hyderbad}\\
Hyderabad, India \\
kamal@iiit.ac.in}
}

\maketitle

\begin{abstract}
Consider a 2D coordinate space with a set of red and a set of blue points. We define a scenic point as a point that is equidistant to a red point and a blue point. The set of contiguous scenic points form a scenic path. The perpendicular bisectors to the line joining a red point and a blue point forms a scenic path between the red point and the blue point. The scenic perpendicular bisectors between different pairs of a red point and a blue point can intersect, forming a scenic graph of intersection points and edges. A scenic route is a traversal on a scenic graph. In this paper, we address this novel problem by (i) designing algorithms for scenic route generation, (ii) studying the algorithms' different properties and (iii) analysing the routes generated by these algorithms. Scenic routes have applications in geo-spatial visualizations and visual analytics.
\end{abstract}


\section{Introduction}
\label{sec:introduction}

\subsection{\textbf{Our Motivation}}
Humans have an implicit awareness of scenic beauty when traveling over various routes. The question arises: How do we transfer our scenic awareness over to two-dimensional data? Here, scenic beauty is defined within the context of a user and a set of viewpoints. Given a context, one can define scenic beauty to be a uniform and balanced view of points of interest. 
In particular, we define a point to be one that provides us with an equidistant view of at least two points of interest with the purpose of achieving balanced views.

Thus, the onus in this work is to show the nature of scenic routes in a two-dimensional environment. We consider points of interest to be either red or blue colored, with scenic beauty being an equidistant view of a red and a blue point. In particular, we introduce scenic routes to pairs of (red, blue) points and give illustrative situations where such routes are viable in the real world. Moreover, we consider the scenic quality of these routes and present different characteristics of the scenic beauty offered by a route.

\begin{figure}
    \centering
    \begin{minipage}{0.48\textwidth}
        \center
        \includegraphics[width=\textwidth]{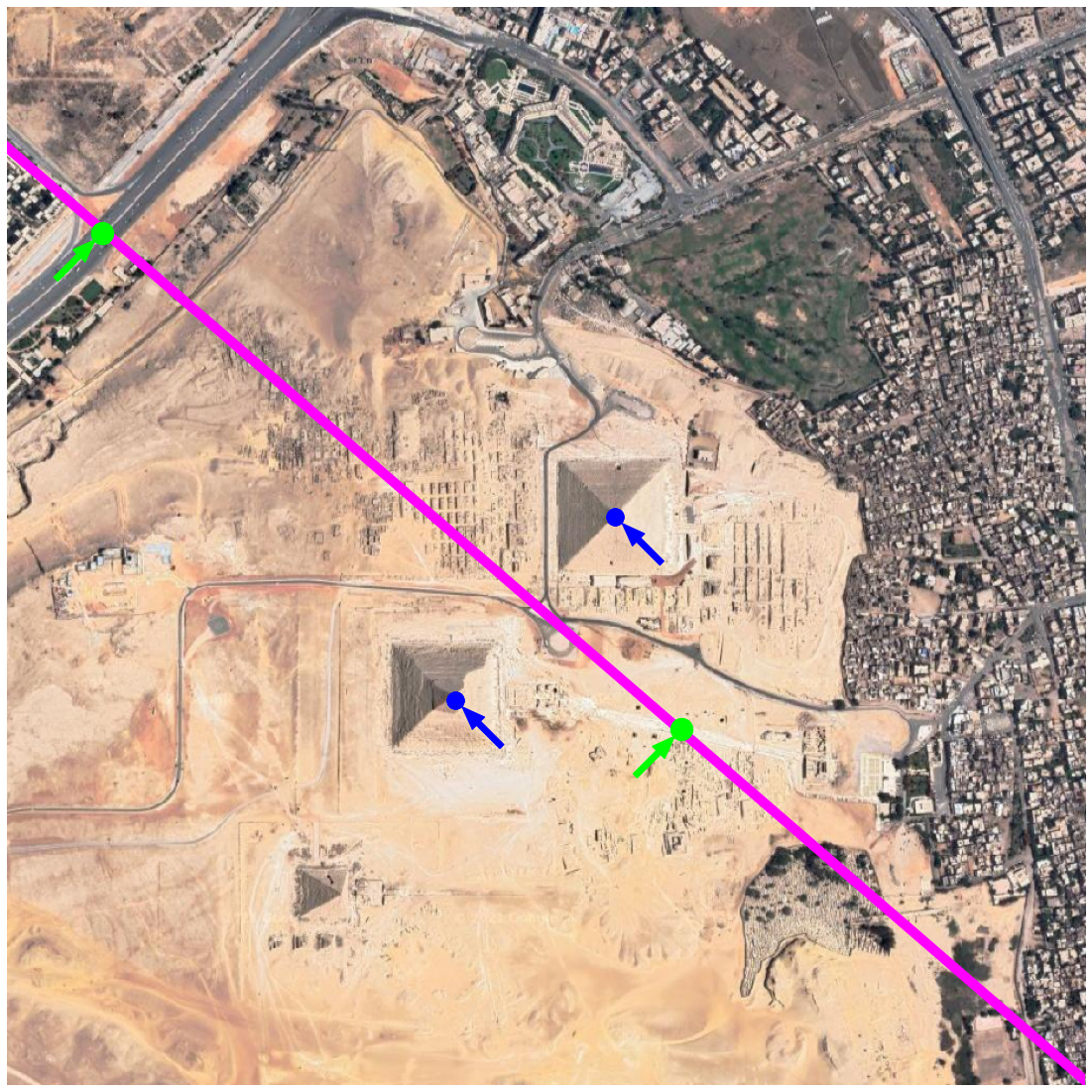} 
        \caption{The Giza Necropolis}
        \label{fig:giza}
    \end{minipage}\hfill
    \begin{minipage}{0.5\textwidth}
        \centering
        \includegraphics[width=\textwidth]{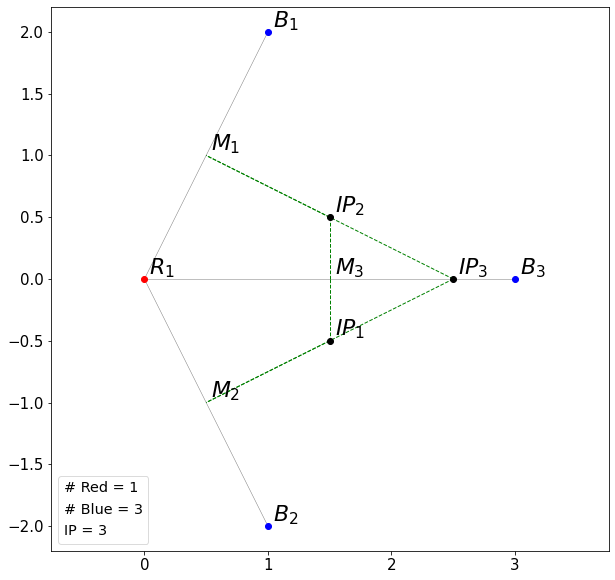} 
        \caption{Red/Blue points, bisector intersection points, and the corresponding scenic graph.}
        \label{fig:start}
    \end{minipage}
\end{figure}

Consider the Giza Necropolis as shown in Fig.~\ref{fig:giza}, specifically the pyramid of Khufu and the pyramid of Khafre. Treating the tops of the two pyramids as points of interest (shown in blue), we can draw a scenic route (shown in pink) through the necropolis. Traveling on this scenic route will guarantee an equidistant view of both points of interest, giving us a balanced (scenic) view of both pyramids. Highlighted in green are two points on the scenic route. The city roads and the scenic route intersect at these points: such points are accessible to the viewer and give a scenic view of the pyramids.



\subsection{\textbf{Scenic Route Problem}}
\label{ssec:problem_intro}
Consider a rectangular region with $N$ red and $M$ blue points. We define a  \bo{scenic point} as a point that is equidistant to a red point and a blue point in a pair of red-blue points. A path on which each point is scenic is termed as a \bo{scenic path}. Likewise, a collection of scenic paths forms a \bo{scenic route}. Each point on the perpendicular bisector of a line joining a pair of red-blue points is scenic. Therefore, the entire perpendicular bisector is a scenic path. Scenic routes are generated by putting together segments of these bisectors/scenic paths. The intersection points of multiple bisectors offer multiple scenic views as they produce an equidistant view to multiple pairs of points. Moreover, these intersection points offer opportunities to change directions, i.e., move from one bisector to another while maintaining a scenic view.

Sets of red and blue points can also be considered as two separate classes in a dataset. In a classification problem, the scenic routes across points from two classes may provide insight regarding the closeness and relation between the classes.

In this paper, we design methods to find scenic routes in 2-dimensional space. Since the routes must be scenic, we can ignore the actual red-blue points and only focus on the bisectors and the points where the bisectors intersect. Moreover, we can transform this set of intersection points and scenic segments between them into a graph connected by the bisectors forming a "scenic graph." A traversal over such a scenic graph would give us a scenic route.



\subsection{\textbf{Formulation}}
\label{ssec:problem_math}
Let the set of red points be $R$, the set of blue points be $B$, and the set of points where the bisectors intersect be $I_P$. Segments of the bisectors between these intersection points form the edge set $E_P$, which contribute as edges of the scenic paths. The weight of each edge is the Cartesian distance between its endpoints. 



The graph over all the intersection points is $G(I_P, E_P)$. Moreover, for a selected route; let the set of points on a route be $S \subseteq I_P$, and the edges within the route be $E \subseteq E_P$. We represent a scenic route as a graph using the notation: $G(S, E)$.

We explain the problem further by using Figure \ref{fig:start}. Consider the point configuration containing one red point $\{R_1\}$ and three blue points $\{B_1, B_2, B_3\}$. $\{M_1, M_2, M_3\}$ represent the midpoints of the lines joining them. The green lines are the perpendicular bisectors of these lines. The green lines represent scenic paths. The intersection points of these perpendicular bisectors are $\{IP_1, IP_2, IP_3\}$. Therefore, to obtain a scenic route, we consider the perpendicular bisectors. Further, to move from one bisector to another, we use their intersection points. In this case:

$$I_P = \{IP_1, IP_2, IP_3\}$$
$$E_P = \{[IP_1, IP_2],\ [IP_2, IP_3],\ [IP_3, IP_1]\}$$

The corresponding scenic graph is $G(I_P, E_P)$. Edges and points within a scenic route are in pink.

\subsection{\textbf{Paper Contribution and Organization}}
\label{ssec:problem_contrib}
The contribution of this paper is: (1) articulation of the scenic routes problem, (2) introduction of scenic routes (based on the equidistant principle) in two-class point configurations in 2D spaces, (3) design scenic route generation algorithms fulfilling different requirements. 

Note that the problem is challenging in 2D space with two sets of points. Increasing the number of dimensions and the number of colors is beyond the scope of this paper.

In Section \ref{sec:routes} we delve into the meaning of scenic routes, lay down requirements for a route to be considered as scenic, and deal with a possible spatial restriction that can be applied to $I_P$. Section \ref{sec:algorithms} contains objectives, explanations, and the motivations behind each scenic route generation algorithm. Section \ref{sec:dicussion} then discusses how well the routes generated using our algorithms perform on the scenic requirements.
\section{Scenic Routes/Traversals}
\label{sec:routes}
A scenic route is defined as a route over $I_P$ via the bisectors (since it is over $I_P$ and we only use bisectors, it is guaranteed to consist of scenic paths). Many such routes with different characteristics are possible. Therefore, there needs to be a characterization of the properties of a \bo{scenic route}. 

\begin{figure}
    \centering
    \begin{minipage}{0.49\textwidth}
        \center
        \includegraphics[width=\linewidth]{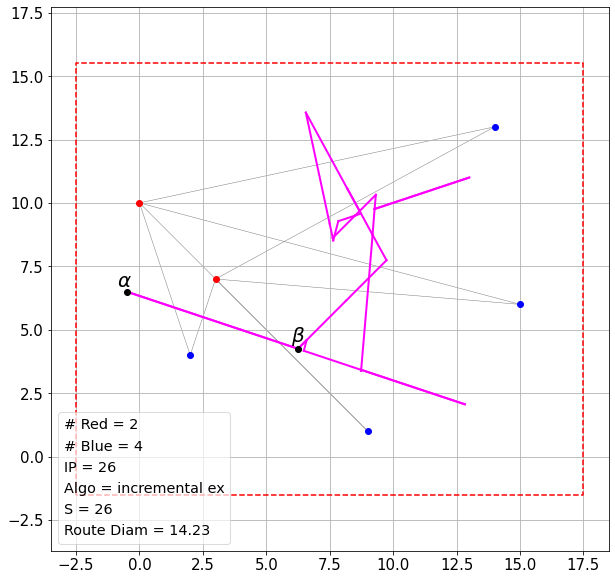}
        \caption{A repeated edge $[\alpha, \beta]$. \textcolor{darkgray}{$(\lvert R\rvert = 2, \lvert B\rvert = 4, \lvert I_P\rvert = 26)$}}
        \label{fig:repeated_edge}
    \end{minipage}\hfill
    \begin{minipage}{0.49\textwidth}
        \center
        \includegraphics[width=\linewidth]{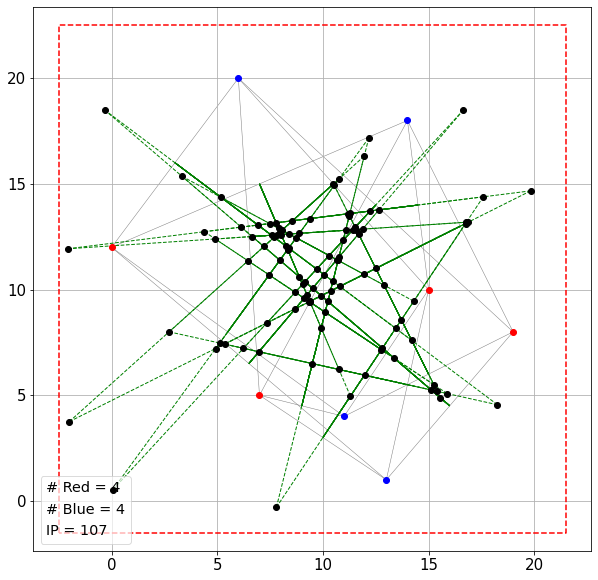}
        \caption{ A configuration restricted by a bounding box. \textcolor{darkgray}{$(\delta = 2.5, \lvert R\rvert = 4, \lvert B\rvert = 4, \lvert I_P\rvert = 107)$} }
        \label{fig:bbox}
    \end{minipage}
\end{figure}

\textbf{A scenic route has the following \underline{requirements}, in decreasing order of importance:}
\begin{enumerate}
    \item \label{routes:scenic} \textbf{Only Scenic}: The route must consist of scenic paths. Any non-scenic path is distracting and must be avoided.
    \item \label{routes:views} \textbf{Completeness}: Travelling on the route must allow one to have a view of a large number of red-blue pairs. It is preferable for a route to give a view of all red-blue pairs (that is, all $|R|\cdot|B|$ pairs). The ability to view a larger number of red-blue pairs on a scenic route would add to the scenic beauty offered by the route. Ideally, all scenic views for all pairs of red-blue points must be covered by a scenic route.
    \item \label{routes:long} \textbf{Minimal Edges}: A route must not have a large number of edges. Traveling on a route must allow for long, uninterrupted stretches of scenic points. In other words, there should not be a large number of direction changes within a route. 
    \item \label{routes:repeated} \textbf{Minimal Repeated Edges}: A route must minimize the number of repeated edges. Repeated edges are defined as stretches of bisectors that must be traveled multiple times (repeated) to complete the entire route. Repeated edges come into play in order to produce a closed path to return to the starting point. Repeated edges within scenic routes offer the same view multiple times. These edges unnecessarily increase the total length of the scenic route without offering any novel scenic views. Hence they should be minimized. In Fig. \ref{fig:repeated_edge}, the edge $[\alpha, \beta]$ is an example of a long repeated edge. If a user was to travel on this edge, the user would also need to travel back on the same edge, i.e., repeat the edge.
\end{enumerate}


To better understand why the above requirements and the ordering of these requirements generates a preferred scenic route, consider a person walking on a scenic route.
\begin{itemize}
    \item At all points during travel, the person should have a scenic view available (Req.~\ref{routes:scenic}).
    \item Given that the red-blue points are points of interest, a person would want to view as many (preferably all) red-blue points as possible (Req.~\ref{routes:views}). 
    \item A person walking on a route would dislike a large number of direction changes, and would prefer longer paths to not distract from the scenic beauty available. (Req.~\ref{routes:long}).
    \item Finally, a person would not want to traverse on the same path repeatedly because repeated edges do not offer any additional novel views, therefore unnecessarily increasing the total distance that needs to be traversed (Req.~\ref{routes:repeated}).
\end{itemize}

A route that fulfills all four requirements listed above is a \textit{preferred} scenic route. A route that fulfills fewer than four requirements is still scenic; however, such a route would not be preferred. For instance, a route may fulfill the first three requirements but include some repeated edges.

Moreover, the importance of each of these requirements and their tradeoffs must be taken into account when generating scenic routes. Having no repeated edges but viewing a small number of red-blue pairs is worse than having a number of repeated edges but viewing a large number of red-blue pairs. 
\subsection{Spatial Restrictions}
\label{sec:spatial}

The perpendicular bisectors between two pairs of red-blue points can be almost parallel, generating intersection points far from the red-blue pairs themselves. While the view given by such points is still scenic according to our definition, it is not a practical view. In such cases, it makes sense to remove such distant intersection points from the set $I_P$.

We remove these intersection points by using a bounding box around the set of red and blue points. The bounding box does not need to be a minimum bounding box; it can be made larger or smaller by some quantity $\delta$. Refer to Fig \ref{fig:bbox} for an example of a configuration restricted by a bounding box.

\section{Scenic Route Algorithms}
\label{sec:algorithms}

As mentioned in Section~\ref{ssec:problem_math}, the problem is to generate scenic routes over the set of bisectors and intersection points. In this section, we design two scenic route generation algorithms and their underlying motivations and objectives.

Each algorithm needs the graph $G(I_P, E_P)$, where $I_P$ is the set of intersection points of the bisectors and $E_P$ is the set of edges (scenic paths) between these intersection points. The algorithms designed require knowing the shortest path that connects a pair of points. This shortest path is computed using the Floyd-Warshall all-pairs shortest path algorithm. We use the notation $d(a,b)$ to denote the distance between two points taken from the precomputed results of the Floyd-Warshall algorithm. The algorithms use the pre-calculated output of the All Pairs Shortest Path(APSP) Algorithm on $G(I_P, E_P)$.

We present the below algorithms that determine scenic routes. We will later analyze the presented algorithms to understand whether they satisfy the requirements of a preferred scenic route. 

\begin{figure}[!t]
    \center
    \includegraphics[width=\columnwidth]{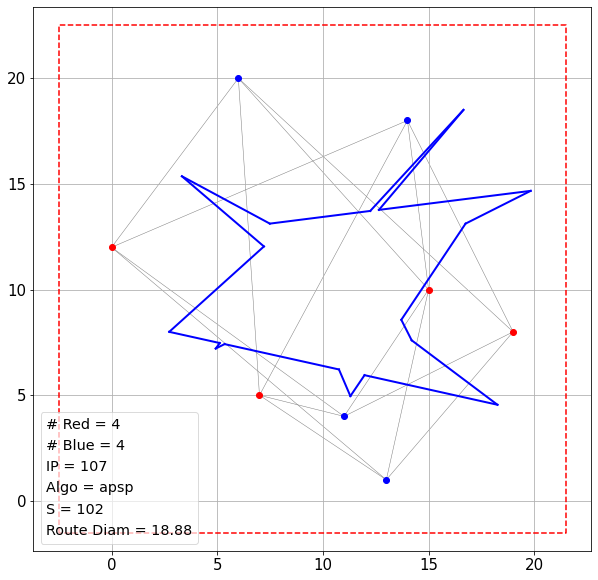}
    \caption{\textbf{Min-Max Hull}, bounded by a bounding box. \textcolor{darkgray}{$(\lvert R\rvert = 4, \lvert B\rvert = 4, \lvert I_P\rvert = 107)$}}
    \label{fig:minmax_apsp}
\end{figure}

\subsection{\bo{Min-Max Hull}}
\label{ssec:minmax_apsp}

The Min-Max Hull algorithm has two parts. First, a Min-Max function selects the intersection points to be considered. Next, the Min-Max Hull Algorithm connects the selected intersection points to create a closed scenic route.

\subsubsection{Min-Max Function:} Given a set of intersection points $I_P$, the Min-Max function focuses on choosing the set of points that should be considered to create a scenic route. The algorithm itself is a refined version of the core idea of the CURE (Clustering Using REpresentatives) \cite{cure} algorithm. 

Let $S$ be the set of points in a scenic route (initially $\emptyset$). The idea behind the Min-Max function is to find new points $x \in I_P-S$ that are neither too close to $S$, nor too far away from it. If we chose points that are too close to $S$, we would end up with cluttered routes and perform worse on Requirement \ref{routes:long} of scenic routes. On the other hand, if we chose points that are too far from $S$, we would end up with routes that go too far from the red-blue points to get meaningful scenic views. The concept of first maximizing the distance and then choosing the closest (minimal distance) point allows us to do precisely that. 

The function finds potential intersection points to be added to the set $S$ where $S$ is the set of intersection points in the output scenic route. Potential intersection points are selected using:
\begin{equation}
    x = \argmin_{x_i \in (I_P-S)}(\max_{a \in S}(d(x_i,a)))
    \label{eqn:minmax}
\end{equation}
where $d(.,.)$ is the all pair shortest path distance between two nodes and $S$ is the set of intersection points in the output scenic route.

\subsubsection{\textbf{Min-Max Hull Algorithm}} The \textbf{objective} of the Min-Max Hull algorithm is to get a convex scenic route with no repeated edges that satisfies the distance bound. Convex scenic routes perform well on two scenic requirements: they have fairly long, uninterrupted views (Req. \ref{routes:long}) and have no repeated edges (Req. \ref{routes:repeated}).

\textbf{Min-Max Initialization:} Let $x_1, x_2$ be two points in the set $I_P$ which have the least distance between them. Then, $S = [x_1, x_2]$.

In this algorithm, we keep an upper bound $B$ on the total route distance to control the length of the routes generated.

The Min-Max  Hull Algorithm first implements the Min-Max initialization described above. Then the algorithm repeatedly implements the following two steps (steps \ref{minmaxhull:1}, \ref{minmaxhull:2}) until the distance bound $B$ is reached:

\begin{algorithm}
	\caption{Min-Max Hull Algorithm} 
	\label{alg:min_max}
	\begin{algorithmic}[1]
	\item \label{minmaxhull:1} Run the Min-Max function to add a new suitable point into the current set of scenic route intersection points $S$. 
    \item \label{minmaxhull:2} Generate the Convex Hull of $S$. Let the sequence of points given by the convex hull be $\Lambda$. Calculate the length of the convex hull as the sum of the APSP distance between every two consecutive points in the sequence $\Lambda$. 
    \item If the length of the convex hull is smaller than the distance bound $B$, go to step \ref{minmaxhull:1}. Else, the scenic route is the convex hull of the set $S$.
	\end{algorithmic} 
\end{algorithm}

We first generate the convex hull on the set of points $S$ using the QuickHull algorithm \cite{quickhull}. We consider the points in the sequence that the QuickHull Algorithm provides us with. However, a bisector that connects two consecutive points computed by the hull may not exist. In such cases, the APSP algorithm is used to find the intermediary nodes (and the edges introduced by them) that connect the two intersection points. On including these extra edges, the final output may not be a convex hull but an altered version of a convex hull, as can be seen in Figure \ref{fig:minmax_apsp}. One can also use the Graham scan algorithm instead of the QuickHull algorithm to incrementally generate the convex hull of $S$.




\begin{figure*}[!t]
    \centering
    \subfloat[\centering Densest lines without hull (plotted in pink). ]{{\includegraphics[width=0.9\columnwidth]{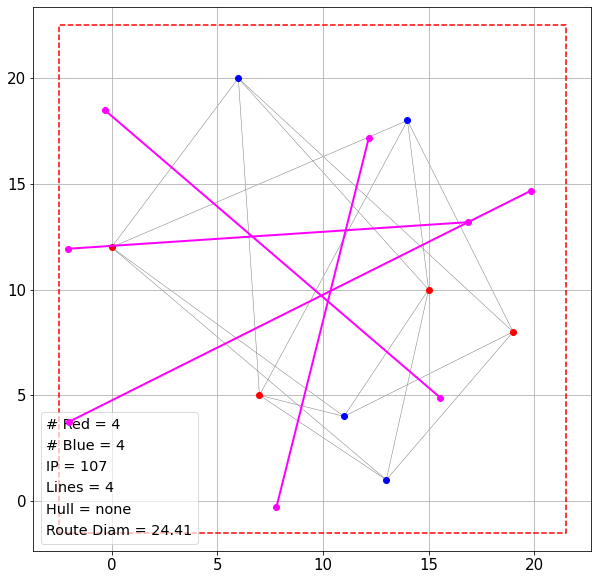} }}
    \quad
    \subfloat[\centering Densest lines with alpha shape hull (hull plotted in dark blue). ]{{\includegraphics[width=0.9\columnwidth]{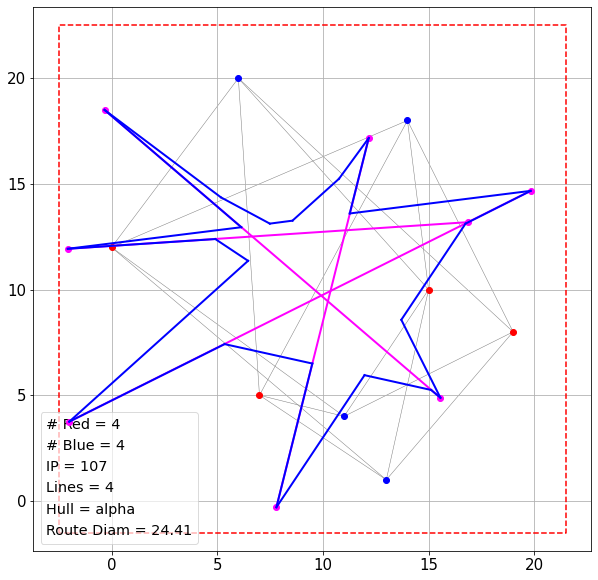} }}
    \caption{\textbf{Densest Line algorithm}, with 4 densest lines highlighted. \textcolor{darkgray}{$(\lvert R\rvert = 4, \lvert B\rvert = 4, \lvert I_P\rvert = 107)$}}
    \label{fig:densest}
\end{figure*}

\subsection{\bo{Densest Line Algorithm}}
\label{ssec:densest_algo}
An intersection point is a point where two bisectors meet. Hence, an intersection point gives scenic views of at least two pairs of red-blue points. A bisector can have several intersecting points on it and hence is a path that represents views to multiple red-blue pairs. The \textbf{objective} of the Densest Line algorithm is to identify such long, straight, uninterrupted scenic paths which reduce directional changes and maximise the number of views.

However, a route solely consisting of long bisectors would perform very poorly on the repeated edges requirement (Req.~\ref{routes:repeated}). Therefore, there need to be some connecting paths between the long bisectors that allow users to view other red-blue pairs without traversing back on the same long bisector. It would make sense to connect the endpoints of the long bisectors using a hull so that once a user reaches the end of a bisector, the user can shift to another bisector using the hull instead of having to backtrack on the same bisector. In particular, we choose alpha shapes \cite{alphashape} as our hull algorithm. 

Alpha Shapes are a generalization of convex hulls over a set of points. While convex hulls are limited to being strictly convex, alpha shapes allow flexibility and can generate concave hulls on a set of points. This can be done by changing the value of the parameter $\alpha$ ($\alpha = 0$ for convex hulls). Specifically, for a point set $S$, the alpha shape of $S$ is the intersection of all closed discs with radius 1/$\alpha$ that contains all the points of $S$.
The alpha shape is generated over the set of endpoints of the long bisectors, thereby connecting the bisectors. The scenic route is comprised of both the dense bisectors and the edges introduced as a result of the alpha shape hull. The user has a choice of whether to continue on the same dense line or shift to another dense line using the edges introduced due to the alpha shape. 

We consider the top $K$ bisectors with the largest number of intersection points. $K$ is a user input mentioning the number of bisectors that need to be considered for creating the scenic route. Increasing $K$ will give more distinct scenic views (Requirement \ref{routes:views}) but will also increase the possible directional changes and also the total route distance.

The Densest Line Algorithm takes the following as input: $K$ representing the number of bisectors to be considered.

\begin{algorithm}
	\caption{Densest Line Algorithm} 
	\label{alg:densest_line}
	\begin{algorithmic}[1]
	\State Rank the bisectors in decreasing order of the number of intersection points.
    \State Pick the top $K$ bisectors from the above ranking. 
    \State \label{step:3_densest}Consider each bisector $L \in K$; let the points $x, y \in I_P$ be the furthest points among all points in $I_P$ that lie on line $L$. Calculate the furthest points for all $K$ bisectors in this manner and let this set of furthest points be $F$.
    \State \label{step:4_densest}Generate the alpha shape (concave hull) \cite{alphashape} of the set $F$. Let the sequence of points given by the alpha shape be $\Lambda$. Connect every two consecutive points in the sequence $\Lambda$ using all pairs shortest path. Note that the points on the shortest path required to connect two consecutive points in $\Lambda$  may not already exist in the set $F$. 
    \State The top $K$ bisectors, alpha shape, and corresponding intersection points are the output scenic route.
	\end{algorithmic} 
\end{algorithm}

Figure \ref{fig:densest} shows an example of the densest line algorithm and the alpha shape hull where the pink lines represent the top $K$ densest lines, and the blue lines are the lines generated on creating the alpha shape as described earlier. 

\begin{figure}
    \centering
    \begin{minipage}{0.49\textwidth}
        \center
        \includegraphics[width=\linewidth]{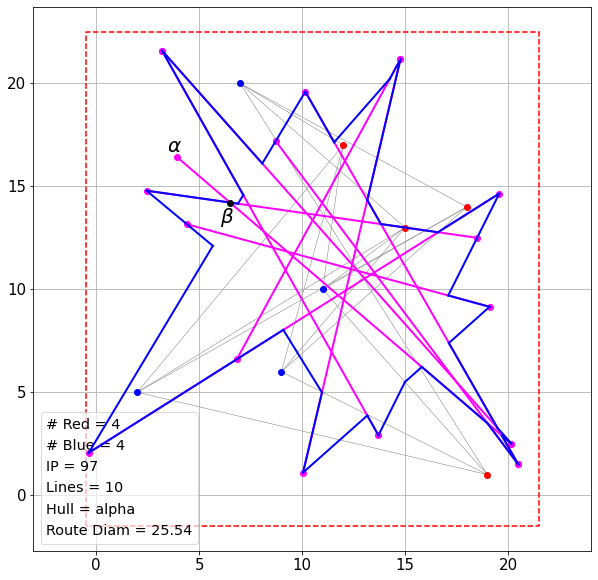}
        \caption{Alpha Shape leading to repeated edges in Densest Line algorithm
        \textcolor{darkgray}{$(\lvert R\rvert = 4, \lvert B\rvert = 4, \lvert I_P\rvert = 97)$}}
        \label{fig:densest_alpha_fail}
    \end{minipage}\hfill
    \begin{minipage}{0.49\textwidth}
        \center
        \includegraphics[width=\linewidth]{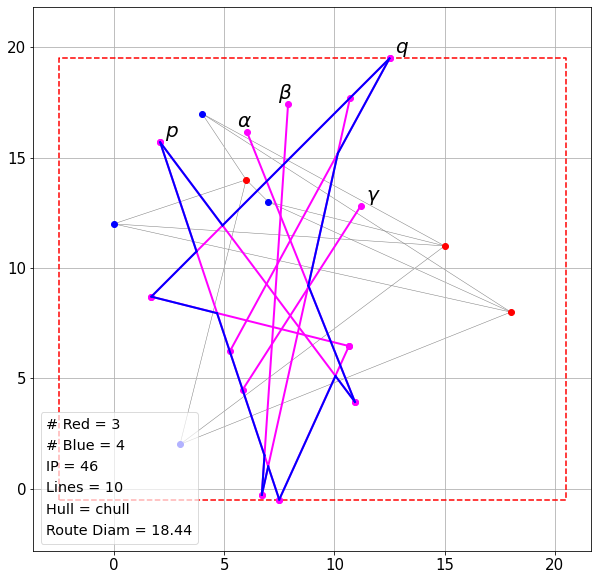}
        \caption{Convex Hull leading to repeated edges in Densest Line algorithm
        \textcolor{darkgray}{$(\lvert R\rvert = 3, \lvert B\rvert = 4, \lvert I_P\rvert = 46)$}}
        \label{fig:densest_chull_fail}
    \end{minipage}
\end{figure}

\textbf{Repeated Edges:} One drawback of using an alpha shape to draw the hull is that there is a possibility of repeated edges occurring. Consider the set $F$ and sequence $\Lambda$ from steps (\ref{step:3_densest}) and (\ref{step:4_densest}) of the algorithm above. Using an alpha shape to get the sequence $\Lambda$ from the set $F$ does not guarantee that each point in $F$ will also be in $\Lambda$. As mentioned earlier, the scenic route is comprised of the points connected using the alpha shape and the $K$ densest lines. The points in $F - \Lambda$ will not be connected using APSP (since they are not in the alpha shape) and may become repeated edges. Refer to Fig.~\ref{fig:densest_alpha_fail}, where edge $[\alpha, \beta]$ is a repeated edge. 

Alpha shapes are preferred over convex hulls for this algorithm because generating a convex hull leads to higher chances of repeated edges occurring. Since we generate the set of points using convex hull algorithms but connect these points using the shortest scenic paths (APSP predecessors) between them, we may miss out on several bisector endpoints. Consider Fig.~\ref{fig:densest_chull_fail}. Due to endpoints $p, q$, the endpoints $\alpha, \beta$ are not included in the convex hull leading them to become repeated edges, as a user traveling from $\beta$ to $\alpha$ has no choice but to travel back to $\beta$ from $\alpha$.



\section{Algorithm Discussion}
\label{sec:dicussion}

As mentioned in Section \ref{sec:routes}, there are requirements for a route to be considered scenic. We now discuss the scenic nature of the routes generated by the algorithms in Section \ref{sec:algorithms}. 

\textbf{Only Scenic (Refer requirement~\ref{routes:scenic}):}
Since both the algorithms generate routes on the bisectors of lines joining red-blue pairs, all the routes generated are compulsorily scenic. Therefore, both Densest Line and Min-Max Hull fulfill this requirement. 

\textbf{Completeness (Refer requirement~\ref{routes:views}):}
A scenic route must give a view of many red-blue pairs (preferably all). For all the algorithms presented, the number of scenic views offered can be tweaked using the user inputs: $B$, the distance bound in the Min-Max Hull algorithm, and $K$, the number of dense bisectors in the Densest Line algorithm.

However, because the number of scenic views is dependent on the user inputs ($K$ and $B$), we cannot make any statements regarding how well scenic routes generated by each algorithm fulfill the Completeness requirement. The completeness requirement will be fulfilled if $K$ and $B$ are sufficiently large. When $K$ is the total number of bisectors or when $B$ is $\infty$, the scenic route will give views of all the red-blue pairs. 


\textbf{Minimal Edges (Refer requirement~\ref{routes:long}):}
It is desirable that a scenic route should have few direction changes and should contain long uninterrupted stretches of scenic views.
\begin{enumerate}
    \item \textbf{Min-Max Hull}: The convex hull contains a large number of fairly straight, long lines. Since the route comprises only the convex hull, the routes contain straight and long lines, leading to few direction changes. 
    \item \textbf{Densest Line}: While the algorithm is motivated by trying to make routes with minimal edges (it picks out long, straight paths in routes), the alpha shape that is plotted on top can give rise to unnecessary direction changes. 
    
\end{enumerate}
While both algorithms perform well on this requirement, the Min-Max Hull algorithm outperforms Densest Line due to the reasons highlighted above.

\textbf{Minimal Repeated Edges (Refer requirement~\ref{routes:repeated}):}
A scenic route has a small number or no repeated edges.

\begin{enumerate}
    \item \textbf{Min-Max Hull}: Because the outer hull is generated using a convex hull, this is a good algorithm for this metric. No repeated edges will exist due to the properties of the convex hull.
    \item \textbf{Densest Line}: Because the alpha shape is used to generate the outer hull, there is a slight possibility of a repeated edge occurring (refer to Fig. \ref{fig:densest_alpha_fail}, point 11 is connected to a repeated edge). In case no outer hull is generated, almost all the edges will be repeated edges (refer to Fig. \ref{fig:densest}). 
\end{enumerate}
Therefore, the Min-Max Hull algorithm (\ref{ssec:minmax_apsp}) outperforms Densest Line on this requirement as well. 

Overall, the \textbf{Densest Line algorithm} is better if one wants to have routes that have long, view-dense stretches, while the \textbf{Min-Max Hull algorithm} is better if one wants to have a shorter route that gives a large number of views.
\section{Illustrations}
\label{sec:illus}

We illustrate our algorithms by generating scenic routes on a real-world example, similar to the illustration in Fig.~\ref{fig:giza}. For these illustrations, we remove the red-blue pairs and instead consider scenic points to be equidistant to \textbf{any two points of interest}. 

\begin{figure}[!t]
    \center
    \includegraphics[width=\linewidth]{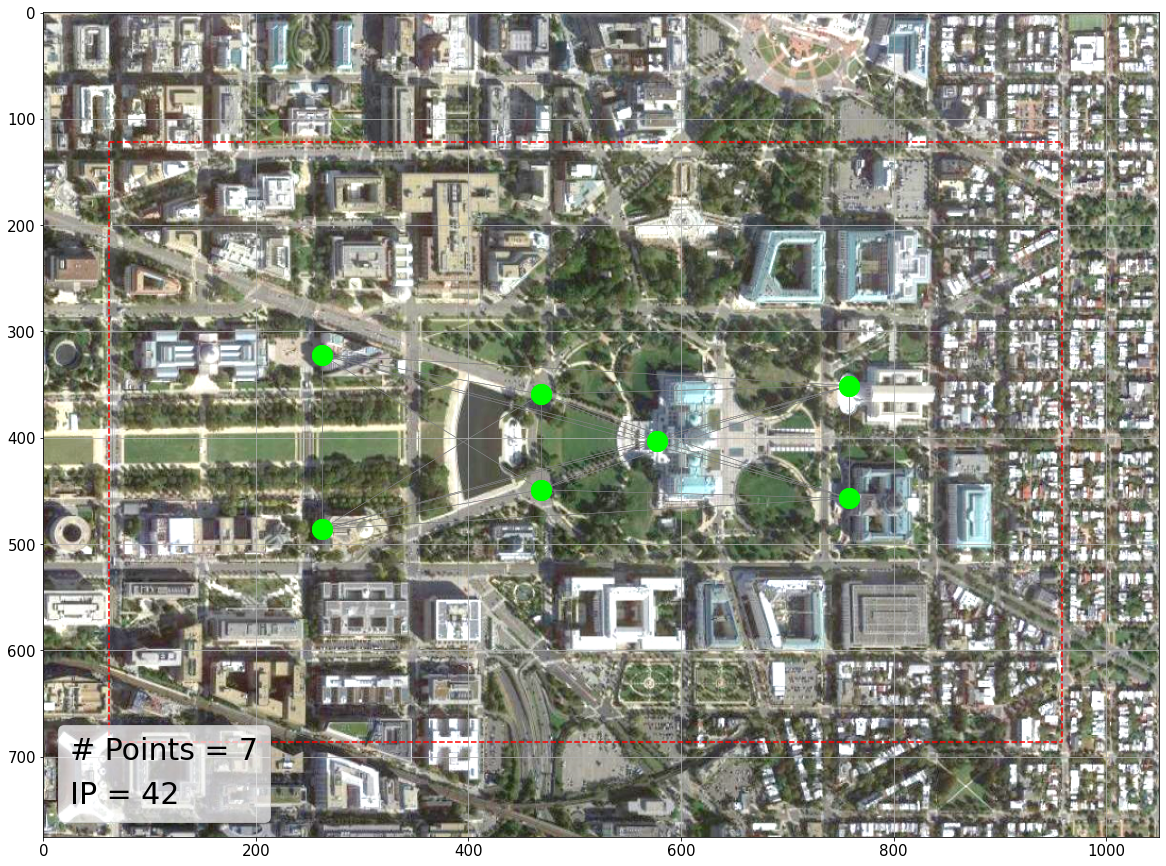}
    \caption{The Capitol Hill area in Washington DC with points of interest highlighted in green}
    \label{fig:capitol}
\end{figure}

The example chosen is the Capitol Hill area in Washington DC, USA. The geographic area along with the chosen points of interest are highlighted (green) in Fig.~\ref{fig:capitol}. The points of interest in the dataset are naturally symmetric, reinforcing the concept of scenic routes based on the equidistant principle. We present scenic routes generated using the Densest Line (Section~\ref{ssec:densest_algo}) algorithm.

\subsection{\textbf{Capitol Hill, USA}}
\label{ssec:illus_capitol}
\begin{figure}[!t]
    \center
    \includegraphics[width=\linewidth]{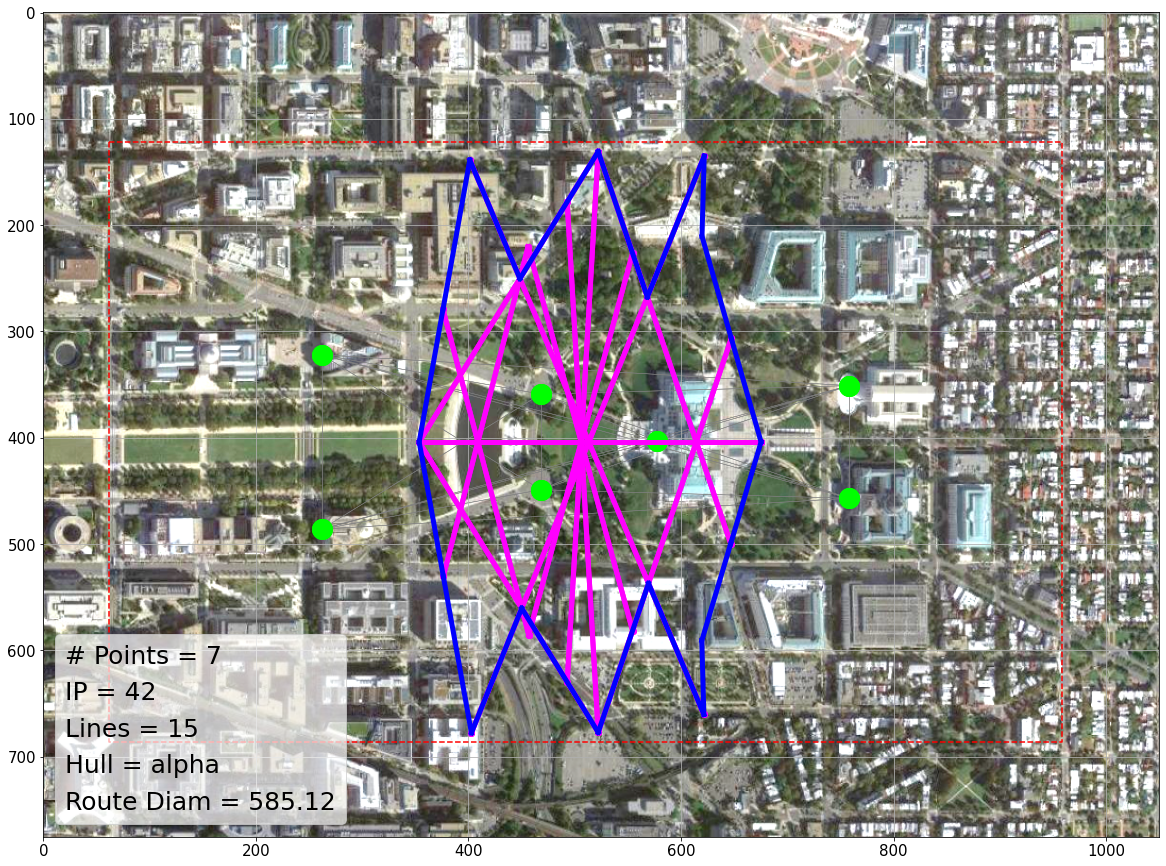}
    \caption{A scenic route on Capitol Hill generated using the densest line algorithm (Refer to Sec.~\ref{ssec:densest_algo}).}
    \label{fig:capitol_densest}
\end{figure}

The seven points of interest chosen are the US Capitol, the Supreme Court, the Library of Congress, the National Museum of the American Indian, the National Gallery of Art, and the Garfield and Peace Monuments. There are $42$ bisector intersection points within the bounding box.
A scenic route generated using the Densest Line algorithm (Refer to Sec.~\ref{ssec:densest_algo}) can be found at Fig.~\ref{fig:capitol_densest}.


\subsection{Discussion}
\label{ssec:illus_discuss}
The Capitol Hill contains city roads that provide routes to the points of interest. The scenic paths in the scenic routes generated by the densest line algorithm match many of the city roads and hence assure the "scenic" aspect of these roads. Other scenic paths intersect city roads at specific points; therefore, those roads give scenic views of points of interest at some locations.


\section{Conclusion}
\label{sec:conclusion}

Given a set of red and blue points in 2D space, a scenic point is equidistant to a red point and a blue point. We develop the problem of scenic points and scenic paths to scenic routes and traversals.

In this paper, we introduced the concept of scenic routes in two-class point configurations in 2D spaces and a characterization of the properties of a scenic route. We present two scenic route generation algorithms and analyze the routes generated by these algorithms. Finally, we generate scenic routes on the Capitol Hill area.

Ours is a preliminary work on this problem that opens up exciting theory and practical implementation challenges to design and analyze scenic route algorithms. The notion of equidistant scenic points provides a balance and beauty to a scenic path. There can be other definitions of scenic points with corresponding scenic routes, but the properties and strategies of densest line and min-max may not change, making them universal approaches to address the problem. 

\bibliography{bibfiles/references}
\bibliographystyle{ACM-Reference-Format}




\end{document}